\documentstyle[prd,aps,preprint]{revtex}
\tightenlines
\input epsf.tex
\def\DESepsf(#1 width #2){\epsfxsize=#2 \epsfbox{#1}}
\begin{document}

\draft
%@twocolumnfalse\endcsname
\preprint{\vbox{
\hbox{OSU-HEP-99-07}}}
\title{Natural Fermion Mass Hierarchy and \\ New Signals for the Higgs
Boson}

\author{{\bf K.S. Babu} and {\bf S. Nandi}}
\address{Department of Physics, Oklahoma State University,
Stillwater, OK, 74078, USA}
\date{June, 1999}
\maketitle
\begin{abstract}
We suggest a novel approach towards resolving the fermion mass hierarchy problem
within the framework of the Standard Model.  It is shown that the observed
masses and mixings can be explained with order one couplings
using successive higher dimensional
operators involving the SM Higgs doublet field.
This scenario predicts flavor--dependent enhancement in the
the Higgs boson coupling to the fermions (by a factor of 3 to the
$b$--quark and $\tau$ and by a factor of $5$ to $\mu$ relative to the SM).  It also
predicts flavor changing $\overline{t}ch^0$ interaction with a strength comparable
to that of $\overline{b}bh^0$.  This opens up a new discovery channel for the Higgs
boson at the upgraded Tevatron and the LHC through $t \rightarrow ch^0$ or $h^0 \rightarrow
\overline{t}c + \overline{c}t$.   Additional tests of the framework include
$D^0-\bar{D^0}$ mixing which is predicted to be near the current experimental limit
and a host of new phenomena associated with flavor physics at the TeV scale.
\end{abstract}
\vskip0.4in

\section{Introduction}

One of the major unresolved puzzles in the Standard Model (SM) is the
observed hierarchy in the masses and mixings of quarks and leptons.  The SM readily
accommodates the measured values, but at the
expense of introducing Yukawa couplings which span in strength from $10^{-6}$ (for the
electron) to $1$ (for the top quark).  An understanding of the small couplings
involved is highly desirable and has been much sought after.

In this Letter we propose a new approach towards resolving the fermion
mass hierarchy puzzle.  We show that an extremely good fit to all the masses and
mixings can be obtained by using higher dimensional operators
involving the relevant fermion fields and
successive powers of the SM Higgs doublet field.
We do not employ any singlet scalar field.
These non--renormalizable
operators will have inverse mass dimensions, but the dimensionless couplings
which multiply them can all be of order one.  Thus the small Yukawa couplings
of the SM will emerge naturally in our approach.

Our proposal differs from most other approaches \cite{other} which address the fermion mass
hierarchy puzzle in one crucial way:  Since we do not utilize vacuum expectation
values (VEVs) of scalar fields that are singlets of SM, the scale of flavor
physics cannot be much above the weak scale.  In fact, from our fits to the fermion
masses and mixings, we find that the scale $M$ where flavor symmetries becomes
manifest, is in the range $(1-2)$ TeV.  While such a scale is high enough to be
consistent with all the experimental constraints, it is within reach
of planned accelerators in the
near future for direct detection.  Had we used VEVs of singlet scalar
fields to explain the mass hierarchy as in most other approaches, the scale of
flavor physics could be (and usually is) near the Planck mass.

In our approach, although the spectrum of the theory below a TeV is that of
the minimal Standard Model, there are a variety of new signals associated with
the Higgs boson that are distinct from the SM.  As noted, the effective Lagrangian involves
higher powers of the SM Higgs doublet field.  As a consequence, the couplings of the
Higgs boson to light fermions are enhanced by a flavor--dependent
numerical factor relative to that of the SM.  These enhancement factors are:
$(3,5,7)$ for ($b,s,d)$ and $(\tau, \mu,e)$; and $(1,3,7)$ for $(t,c,u)$.
These numbers are related to the dimension of the operators from which a particular
fermion receives its mass.  The enhanced $\bar{b}bh^0$ coupling (by a factor 3)
would have a significant impact on Higgs detection at hadron colliders; the
enhancement in the muonic coupling implies that the Higgs production cross section
at a muon collider \cite{muon} will be increased by a factor of 25.

A further consequence of using successive higher dimensional operators
is that there are flavor changing neutral current processes
mediated by the SM Higgs boson.  The mass matrix and the Yukawa coupling
matrix do not diagonalize simultaneously, although there is only a single
Higgs doublet field.  We show that all the low energy constraints, such as from
$K^0-\bar{K^0}$ mass difference, are well satisfied.  The model predicts
$D^0-\bar{D^0}$ mixing to be close to the present experimental limit.  An
improvement by a factor of $(10-20)$ could test the model, if the Higgs
boson mass is not much above $300$ GeV.

The proposed scenario predicts an interesting new discovery channel for the Higgs boson
at the upgraded Tevatron and the LHC.  It uses the flavor changing Higgs vertex $\bar{t}ch^0$
which has a strength  comparable to that of the
flavor conserving $\bar{b}bh^0$ vertex.
If the mass of $h^0$ is below 150 GeV, then in the $\bar{t}t$ production at colliders,
the decay of the top quark
can lead to Higgs boson signals with $h^0 \rightarrow \bar{b}b$ providing the
invariant mass of the Higgs while the $W+\bar{b}$ decay of the $\bar{t}$ tagging the event.
If the Higgs mass is above $m_t$ ($m_{h^0} \approx
(200-350)$ GeV), the production
process $gg \rightarrow h^0$ followed by $h^0 \rightarrow \bar{t}c+\bar{c}t$
can lead to a new signal at LHC.

In the following we will describe in more detail the scheme and its predictions
just outlined.  We will also comment on some possible ways of realizing the effective
Yukawa Lagrangian from renormalizable theories with flavor symmetries broken at the TeV scale.

\section{Fermion mass hierarchy from higher dimensional operators with SM Higgs}

We assume that some flavor symmetries prevent the direct Yukawa coupling of the SM Higgs doublet
($H$) to the light fermions.  These flavor symmetries are spontaneously broken at a scale
$M$, which will be seen to be in the $(1-2)$ TeV range.  The effective theory below
$M$ is the SM with one Higgs doublet but with non--renormalizable terms in the
Higgs Yukawa couplings.  For the quark sector the effective Yukawa Lagrangin
is taken to be (in an obvious notation for the quark fields and with
$\tilde{H} \equiv -i \tau_2 H^*$):
\begin{eqnarray}
{\cal L}^{\rm Yuk} &=& h_{33}^u Q_3 u_3^c \tilde{H} + \left({H^\dagger H \over M^2}\right)
\left(h_{33}^d Q_3 d_3^c H + h_{22}^u Q_2 u^c_2 \tilde{H}+h_{23}^u Q_2 u_3^c \tilde{H}
+ h_{32}^u Q_3 u_2^c \tilde{H}\right) \nonumber \\
&+& \left({H^\dagger H \over M^2}\right)^2
\left(h_{22}^dQ_2 d_2^c H + h_{23}^d Q_2 d_3^c H + h_{32}^d Q_3 d_2^c H+ h_{12}^uQ_1u_2^c
\tilde{H} + h_{21}^uQ_2u_1^c \tilde{H} \right. ) \nonumber \\
&+& \left.  h_{13}^uQ_1 u_3^c \tilde{H} + h_{31}^u Q_3 u_1^c
\tilde{H} \right)
+ \left({H^\dagger H \over M^2}\right)^3 \left(h_{11}^u Q_1 u_1^c \tilde{H}
+ h_{11}^dQ_1 d_1^c H  \right.  \nonumber \\
&+& \left. h_{12}^dQ_1 d_2^c H + h_{21}^d Q_2 d_1^c H + h_{13}^d Q_1 d_3^c H
+ h_{31}^d Q_3 d_1^c H \right ) + H.C.
\end{eqnarray}
Only the top quark has a renormalizable Yukawa coupling, other couplings are
suppressed by successive powers of $(H^\dagger H/M^2)$.
This provides the
small expansion parameter needed to explain the light fermion masses.  We will show that
an good fit to all the fermion masses and mixing angles can be obtained
with the dimensionless couplings ($h_{ij}^{u,d}$) in Eq. (1) taking values
of order one.  Note that $SU(2)_L$ invariance
prevents even number of Higgs doublet fields in the expansion.  In each term
of Eq. (1) there is a unique $SU(2)$ group contraction.

In most other attempts to explain the mass hierarchy, singlet scalar
fields ($S_i$) are employed and the expansion parameter is
$\left\langle S_i \right \rangle/M$.  In such cases, both $\left\langle S_i
\right \rangle$ and $M$ both can be large, for eg., near the Planck scale.  As long as
$\left\langle S_i\right\rangle/M$ is finite (i.e., not extremely small), a fit
to the fermion masses can be obtained.
There is no a priori
reason in such cases
for the flavor symmetry to be broken near the weak scale.  The low energy
theory will be identical to the SM with no modification in the Higgs boson
interactions.  In contrast, in our
approach, since no singlet fields are employed,
the expansion parameter is $\left\langle H^\dagger H \right\rangle/M^2$,
which implies that $M$ cannot be much above the weak scale.  It also results
in new interactions of the SM Higgs boson, which can be directly tested.

We will comment on possible ways of deriving the effective Lagrangian of Eq. (1)
from renormalizable theories towards the end of the paper.  We remark here
that the coefficients $h_{ij}^{u,d}/M^2$ could be thought of as background fields
which carry flavor quantum numbers.  In writing Eq. (1), we have allowed for
the possibility that the underlying flavor structure might be parity invariant.
Identify $h_{ij} = h_{ji}^*$ in that case.  The  dimension of the
various operators in Eq. (1) are
determined by examining the magnitudes of the
quark masses and mixings.  We have allowed for this to be as low as
it can be, consistent with our demand that the dimensionless coefficients be
order one.  This is the most general case.
Interesting special cases which might arise due to the constraints of
flavor symmetry such
as $h_{13}^{u,d}=h_{31}^{u,d} = 0$ or $h_{ij}^d = 0 (i \neq j),$ can be readily
incorporated into our analysis.  The effective Lagrangian inducing the charged lepton
masses is taken to have a form identical to that for the down--type quarks (replace
the couplings $h_{ij}^d$ by $h_{ij}^\ell$ in Eq. (1) for the leptons).

Writing $H = (h^0/\sqrt{2}+v,~~0)^T$ in unitary gauge with $v \simeq 174$ GeV,
and defining a small parameter
\begin{equation}
\epsilon \equiv { v \over M},
\end{equation}
Eq. (1) leads to the following mass matrices for the up--type and the down--type
quarks:
\begin{eqnarray}
M_u = \left(\matrix{h_{11}^u \epsilon^6 & h_{12}^u \epsilon^4 & h_{13}^u \epsilon^4 \cr
h_{21}^u\epsilon^4 & h_{22}^u \epsilon^2 & h_{23}^u \epsilon^2 \cr
h_{31}^u \epsilon^4 & h_{32}^u \epsilon^2 & h_{33}^u}\right)v, ~~~~~
M_d = \left(\matrix{h_{11}^d \epsilon^6 & h_{12}^d \epsilon^6 & h_{13}^d \epsilon^6 \cr
h_{21}^d\epsilon^6 & h_{22}^d \epsilon^4 & h_{23}^d \epsilon^4 \cr
h_{31}^d \epsilon^6 & h_{32}^d \epsilon^4 & h_{33}^d \epsilon^2}\right)v~.
\end{eqnarray}
The charged lepton mass matrix is obtained from $M_d$ by replacing
$h_{ij}^d \rightarrow h_{ij}^\ell$.  The couplings $h_{ij}$ are complex
in general, but $\epsilon$ can be taken to be real without loss of generality.

The masses of the quarks and leptons can be read off from Eq. (3) in the
approximation $\epsilon \ll 1$.  They are:
\begin{eqnarray}
\{m_t, m_c, m_u\}  &\simeq& \{|h^u_{33}|,~ |h^u_{22}|\epsilon^2,~ |h^u_{11} -
h^u_{12}h_{21}^u/h_{22}^u|\epsilon^6\}v ,\nonumber \\
\{m_b, m_s, m_d\} &\simeq& \{|h_{33}^d| \epsilon^2, ~|h_{22}^d| \epsilon^4,~
|h_{11}^d|\epsilon^6\}v, \nonumber \\
\{m_\tau, m_\mu, m_e\} &\simeq& \{|h_{33}^\ell| \epsilon^2,~ |h_{22}^\ell| \epsilon^4,~
|h_{11}^\ell|\epsilon^6\}v.
\end{eqnarray}
The quark mixing angles are found to be:
\begin{eqnarray}
|V_{us}| &\simeq& \left|{h_{12}^d \over h_{22}^d}-{h_{12}^u \over h_{22}^u}\right| \epsilon^2,
\nonumber \\
|V_{cb}| &\simeq& \left|{h_{23}^d \over h_{33}^d}-{h_{23}^u \over h_{33}^u}\right|
\epsilon^2 ,\nonumber \\
|V_{ub}| &\simeq& \left|{h_{13}^d \over h_{33}^d}-{h_{12}^u h_{23}^d \over h_{22}^uh_{33}^d}
- {h_{13}^u \over h_{33}^u}\right|\epsilon^4.
\end{eqnarray}

In order to see how well our scheme explains the observed
mass and mixing hierarchy, we choose an illustrative
set of input values for the quark masses \cite{gasser}: $m_u(1~{\rm GeV}) = 5.1~
{\rm MeV},~m_c(m_c) = 1.27~{\rm GeV},~ m_t^{\rm phys} = 175~{\rm GeV},~m_d(1~{\rm GeV})
= 8.9~{\rm MeV}, ~m_s(1~{\rm GeV}) = 175~{\rm MeV},~m_b(m_b) = 4.25~{\rm GeV}$, along
with the well known charged lepton masses.  We extrapolate all masses to a common
scale, conveniently chosen as $m_t(m_t) \simeq 166$ GeV using three loop QCD and
one loop QED beta functions.  With $\alpha_s(M_Z) = 0.118$, the values of the running
masses at $\mu = m_t(m_t)$ are found to be:
\begin{eqnarray}
\{m_t, m_c, m_u\} &\simeq& \{166,~ 0.60,~ 2.2 \times 10^{-3}\}~{\rm GeV}, \nonumber \\
\{m_b, m_s, m_d\} &\simeq& \{2.78,~ 7.5 \times 10^{-2},~ 3.8 \times 10^{-3}\}~{\rm GeV},
\nonumber \\
\{m_\tau, m_\mu, m_e\} &\simeq& \{1.75,~ 0.104,~ 5.01 \times 10^{-4}\}~{\rm GeV}.
\end{eqnarray}
An optimal choice of the expansion parameter is $\epsilon = 1/6.5$.  ($\epsilon \simeq
1/7-1/6$ gives reasonable fits.)
This corresponds to a relatively low scale of flavor symmetry breaking: $M \simeq 1.1$
TeV.  With $\epsilon = 1/6.5$, we determine the dimensionless coefficients $h_{ij}^{u,d,\ell}$
from the fermion masses quoted in Eq. (6).  They are:
\begin{eqnarray}
\{|h_{33}^u|, |h_{22}^u|, |h_{11}^u-h_{12}^u h_{21}^u/h_{22}^u|\} &\simeq& \{0.96, 0.14,
0.95\}, \nonumber \\
\{|h_{33}^d|, |h_{22}^d|, |h_{11}^d|\} &\simeq& \{0.68, 0.77, 1.65\}, \nonumber \\
\{|h_{33}^\ell|, |h_{22}^\ell|, |h_{11}^\ell|\} &\simeq& \{0.42, 1.06, 0.21\}.
\end{eqnarray}
We see that, remarkably, all couplings are of order unity.  The largest deviation
from one is the charm Yukawa coupling $|h_{22}^u| \simeq 0.14$.  This small
fluctuation actually goes in the right direction to explain the magnitude of the
Cabibbo angle.  From the expression for $|V_{us}|$ (Eq. (5)), with $|h_{22}^u| \simeq
0.14$, and $h_{12}^{u,d}$ of order one, the correct value of
$|V_{us}|\simeq 0.2$ follows naturally.  (In our scheme, $|V_{us}|$ is a linear
combination of
${\cal O}(\sqrt{m_u/m_c})$ and  ${\cal O} (m_d/m_s)$ terms.)
Similarly, $|V_{ub}| \sim 7 \epsilon^4 \simeq 0.004$, where
the factor $7$ is due to $1/h_{22}^u$ enhancement in
the second term of $|V_{ub}|$ (see Eq. (5)).
$|V_{cb}| \simeq 0.04$ can be fit for example, by choosing
$|h_{23}^u| \simeq 1.4$ (if that term dominates) or with $|h_{23}^d| \simeq 0.84$
(if it dominates).  We thus see that the fit to all the quark and lepton masses
as well as the mixing angles is
extremely good with all couplings taking order one values.
We regard the success of the
fit to be a strong enough motivation to take
the proposed scheme  seriously and study its
experimental consequences.  This is what we do in the discussions that follow.

\section{Enhanced Yukawa couplings and Higgs mediated FCNC}

The Yukawa coupling matrices of the SM Higgs boson to the up--type and
the down--type quarks that follow from Eq. (1) are:
\begin{eqnarray}
Y_u = \left(\matrix{7h_{11}^u \epsilon^6 & 5h_{12}^u \epsilon^4 & 5h_{13}^u \epsilon^4 \cr
5h_{21}^u\epsilon^4 & 3h_{22}^u \epsilon^2 & 3h_{23}^u \epsilon^2 \cr
5h_{31}^u \epsilon^4 & 3h_{32}^u \epsilon^2 & h_{33}^u}\right), ~~~~~
Y_d = \left(\matrix{7h_{11}^d \epsilon^6 & 7h_{12}^d \epsilon^6 & 7h_{13}^d \epsilon^6 \cr
7h_{21}^d\epsilon^6 & 5h_{22}^d \epsilon^4 & 5h_{23}^d \epsilon^4 \cr
7h_{31}^d \epsilon^6 & 5h_{32}^d \epsilon^4 & 3h_{33}^d \epsilon^2}\right),
\end{eqnarray}
with the charged lepton Yukawa coupling matrix $Y_\ell$ obtained from $Y_d$
by replaing $h_{ij}^d \rightarrow h_{ij}^\ell$.

There are two striking features in Eq. (8).  One is that the diagonal couplings
are enhanced relative to the respective SM Higgs boson Yukawa couplings by a numerical factor.
These enhancement factors are $(1,3,5)$ for $(t,c,u)$ and $(3,5,7)$ for
$(b,s,d)$ as well as $(\tau, \mu, e)$.  In the usual SM, these factors are all
equal to one.  In two Higgs doublet models or in supersymmetric models, while
the couplings of $h^0$ might be enhanced, they are not flavor dependent, nor do they
take these specific values.
The use of higher dimensional operators in our approach
make them flavor--dependent and different from one.  The second feature is related
to the flavor dependence which has
an important consequence: The Yukawa coupling matrix and the corresponding
mass matrix do not diagonalize simultaneously.  This will lead to
Higgs mediated flavor changing neutral current processes, even though there
is only a single electroweak Higgs boson in the theory \cite{2higgs}.

The enhanced fermionic Yukawa couplings has a strong influence on the strategy for discovering
the Higgs boson.  We will discuss this in the next Section.  First we will address
the FCNC processes and establish that they are consistent with low energy
data such as $K^0-\bar{K^0}$ mass difference.

The FCNC couplings of the SM Higgs boson $h^0$ to fermions can be obtained from Eqs. (8) and
(3).  In the quark sector in terms of the mass eigenstates they are given by:
\begin{eqnarray}
{\cal L}^{\rm FCNC} &\simeq& {h^0 \over \sqrt{2}}(2h_{12}^d \epsilon^6 ds^c+
2h_{21}^d \epsilon^6 sd^c+4h_{13}^d \epsilon^6
db^c + 4h_{31}^d \epsilon^6 bd^c + 2h_{23}^s \epsilon^4 sb^c+2h_{32}^d \epsilon^4 bs^c)
\nonumber \\
&+& {h^0 \over \sqrt{2}}(2h_{12}^u \epsilon^4 uc^c+
2h_{21}^u \epsilon^4 cu^c+4h_{13}^d \epsilon^4
ut^c + 4h_{31}^u \epsilon^4 tu^c + 2h_{23}^u \epsilon^2 ct^c+2h_{32}^u \epsilon^2 tc^c) + H.C.
\end{eqnarray}
For FCNC Higgs couplings in
the charged lepton sectors, replace $h_{ij}^d \rightarrow h_{ij}^{\ell}$
in Eq. (9).
The factor in the $ds^c$ vertex is obtained as $7-5=2$, where $5$ is the
coefficient of the (2,2) entry in $Y_d$ and $7$ that of the (1,2) entry.

There is a tree--level contribution mediated by the Higgs boson for $K^0-\bar{K^0}$
mass difference.  We estimate it in the vacuum saturation approximation for the
hadronic matrix element \cite{soni}.
The new contribution, $\Delta m_K^{\rm Higgs}$, is given by
\begin{eqnarray}
\Delta m_K^{\rm Higgs} &\simeq& {4 \over 3} {f_K^2 m_K B_K \over m_{h^0}^2}\epsilon^{12}[
\{ {1 \over 6} {m_K^2 \over (m_d+m_s)^2} + {1 \over 6}\} {\rm Re}\left[\left({h_{12}^d+
h_{21}^{d*} \over \sqrt{2}}\right)^2\right] \nonumber \\
&-& \{ {11 \over 6} {m_K^2 \over (m_d+m_s)^2} + {1 \over 6}
\} {\rm Re}\left[\left({h_{21}^d-h_{12}^{d*} \over \sqrt{2}}\right)^2\right]]~.
\end{eqnarray}
Here $B_K$ is the bag parameter.
The first term arises from the scalar piece of the matrix element, while
the second term is from the pseudoscalar piece.  Note that the pseudoscalar
contribution is larger than the scalar one by about an order of magnitude.  On the
other hand, if the underlying theory is parity invariant, $h_{12}^d = h_{21}^{d*}$,
so only the scalar piece will be significant.  We shall also allow for this possibility.
Using $B_K = 0.75, f_K \simeq 160$ MeV, $\epsilon \simeq 1/6.5$ $m_s = 175$ MeV,
$m_d = 8.9$ MeV,
and with a parity non--invariant choice $h_{12}^d = 1,  h_{21}^d = 0.5$,
we obtain $\Delta m_K^{\rm Higgs} \simeq 3.1 \times 10^{-17}$ GeV, for $m_{h^0} = 100$ GeV.
This is two orders of magnitude below the experimental value.  If parity invariance is
assumed, with $h_{12}^d \simeq h_{21}^{d*} = 1$, the same choice of parameters
gives $\Delta m_K^{\rm Higgs} \simeq 6 \times 10^{-16}$ GeV, which is a factor of 6 below the
experimental value.  We see broad consistency with
data.

As for the CP violating parameter $\epsilon_K$, it receives a
new contribution from the Higgs exchange which can be significant and can even dominate over the
the SM CKM contribution.  For example, in the choice of parameters with
$|h_{12}^d| = 1, |h_{21}^d| = 0.5$, but with their phases being of order
one, $\epsilon_K$ arising from the Higgs exchange can explain the
observed value entirely.  This possibility will be tested at the $B$ factory.
New contributions to $\epsilon'$ are negligible.
However, since the prediction for $\epsilon_K$ is modified, the
standard model fit to the CKM parameter
Im($V_{td}^*V_{ts})$ will be modified by a factor of order one (depending on
the relative strength of the CKM and the Higgs exchange contribution to $\epsilon_K$).
Since $\epsilon'$ is directly proportional to Im($V_{td}^* V_{ts})$, its prediction
will be altered by a factor of order one \cite{buras}.

Electric dipole moments (EDM) of the neutron and the electron in our scheme is much
larger than the SM prediction.  The dominant source of the neutron EDM ($d_n$)
is from the two--loop
Barr-Zee \cite{barr} diagram involving the SM Higgs boson and the $Z$ boson.
In our scheme, $h^0$ has a scalar as well as a pseudoscalar coupling to the
$u$--quark.  In the physical basis of the $u$--quark the pseudoscalar coupling
has a strength of order Im($h_{11}^u) (9 \epsilon^6)$.  From this, we estimate the neutron
EDM to be in the range $(10^{-26}-10^{-27})$ ecm for the phase of order unity
and Higgs mass of order 100 GeV.  The EDM of the electron ($d_e$) will arise from an
analogous diagram, but its magnitude is suppressed by an additional power of
$\epsilon^2$.  We estimate $d_e \approx 10^{-27}-10^{-28})$ ecm.  Both $d_e$
and $d_n$ are within reach of future experiments.

All other constraints from low energy flavor changing processes are easily
satisfied in the model.  For example, the process $K_L \rightarrow \mu^+ \mu^-$
will proceed via tree--level Higgs exchange.  The four--Fermion interaction
for the decay is given by:
\begin{equation}
{\cal L}^{\rm eff} \simeq {5 \over 2} h_{22}^\ell (h_{21}^d-h_{12}^{d*})
{\epsilon^{10} \over m_{h^0}^2} (\bar{d}\gamma_5 s)(\bar{\mu}\mu)+ H.C.
\end{equation}
This amounts to an amplitude of strength $\sim 10^{-7}G_F$,
which implies that the branching ratio for
$K_L \rightarrow \mu^+\mu^-$ will be of order $10^{-14}$ from Higgs exchange.
This is well below the experimental limit.  Other processes such as
$K_L \rightarrow \mu e$, $K \rightarrow \pi \bar{\nu} \nu$, $\mu \rightarrow
e \gamma$, $\mu \rightarrow 3e$, $B_d-\bar{B_d}$ mixing, etc are all orders of magnitude
below the corresponding experimental limits.

$D^0-\bar{D^0}$ mixing, on the other hand, is predicted to be near the present
experimental limit in our scenario.  Note that the FCNC $uc^ch^0$ vertex is
enhanced by a factor of $\epsilon^2$ compared to $ds^ch^0$ vertex (see Eq. (9)).
The new contribution  $\Delta m_D^{\rm Higgs}$ is given by an expression
analogous to Eq. (10).  Using $f_D \simeq 200$ MeV, $B_D \simeq 0.75$,
$m_D^2/(m_c+m_u)^2 \simeq 2$, and $h_{12}^u =1, h_{12}^u = 0.5$, we estimate
$\Delta m_D^{\rm Higgs} \simeq 4.7 \times 10^{-14}$ GeV for a Higgs mass of 200
GeV, to be compared with the present experimental limit of $ \Delta m_D \le
1.6 \times 10^{-13}$
GeV \cite{pdg}.  Even if parity invariance is assumed, the new contribution does not
diminish for the $D^0$ system.  Allowing for reasonable order one uncertainties in the hadronic
matrix element and the FCNC couplings, and varying $\epsilon$ in the
range $1/6-1/7)$, we conclude that $D^0-\bar{D^0}$ mixing
should be not more than a factor of $(10-20)$ below the present limit, provided that
the Higgs boson mass is below about 300 GeV.  This prediction should be
testable in the near future.  We should remark that the SM long distance contribution
to $D^0-\bar{D^0}$ mixing \cite{pakvasa} is expected to be about three orders of magnitude below
the current limit, so its discovery should be clear--cut signal for new
short--distance physics, such as the one we propose.

\section{New signals for the Higgs boson}

There are a variety of new signals associated with the production and decay
of the Higgs boson in our scenario.  While the tree-level Higgs couplings
to the gauge bosons are identical to that of the SM, the Yukawa interactions
are modified from that of the SM.  The consequences are significant for the
strategy to discover the Higgs boson at colliders \cite{higgs}.  We list
below  the processes
where the differences from the SM are most striking.

1.  The Higgs boson couplings to light fermions are enhanced by
a flavor--dependent numerical factor.  At the NLC and perhaps the LHC, the Yukawa couplings
to $(b,\tau,t)$ will be measured \cite{dawson}.  The scenario can thereby be directly
tested.  The enhanced coupling to $\mu$ implies that
the Higgs production cross section will increase by a factor of 25 at
a muon collider \cite{muon}, relative to that of SM.

2.  The branching fraction
for Higgs decay into $\bar{b}b$ will increase by a factor of 9 relative to SM.
In the SM, the $\bar{b}b$ and the $W^*W^*$ branching ratios become
comparable for a Higgs mass of about 135 GeV \cite{spira}.  Because of the difference
in its $b$--quark coupling, this cross--over occurs in our scheme
for $m_{h^0} \simeq 155$ GeV.  This difference should be incorporated into the strategy
for discovering the Higgs boson at the upgraded Tevatron \cite{han}.

3.  The $h^0\gamma\gamma$ coupling, which arises primarily from the
$W^\pm$ loop, is unaffected in our scheme.  Since the $\bar{b}b h^0$
vertex is enhanced by a factor of $9$, the branching ratio for $h^0
\rightarrow \gamma \gamma$ will decrease by a factor of approximately 9, if
the Higgs boson mass is below about 155 GeV.

3.  At hadron colliders such as the upgraded Tevatron and the LHC,
gluon fusion is a dominant source for Higgs production.  Since the
top--quark coupling $\bar{t}th^0$ in our scheme is identical to
that of the SM, the production cross section via gluon fusion is
little affected.  However, there is some difference since the
$b$--quark contribution is not negligible. In the SM, for Higgs
mass of $(100,150,200)$ GeV, the $b$--quark loop contribution to
$gg \rightarrow h^0$ decreases the production cross section by
about $(9,5,3)\%$ (for $m_b(m_b) = 4.25$ GeV). In our scenario,
Higgs production production is decreased from the SM value by
$(17,10,6)\%$.  This should be measurable at
the LHC.

4.  There is an exciting new discovery channel for the Higgs boson at the
upgraded Tevatron.  This utilizes the flavor--changing $\bar{t}ch^0$ vertex.
Note that in our scenario this vertex has a strength of order $2\epsilon^2$,
which is similar in magnitude to that of the flavor conserving $\bar{b}bh^0$
coupling.

If $m_{h^0}$ is less than $m_t$, the decays $t \rightarrow ch^0$
and $\bar{t} \rightarrow \bar{c}h^0$ can provide a new channel for
Higgs discovery.  Once produced, $h^0$ will decay into $\bar{b}b$ with
a significant branching ratio.  The width for $t \rightarrow c h^0$ in our scheme is
given by:
\begin{equation}
\Gamma(t c h^0) \simeq {G_F m_t^3 \over 8\pi \sqrt{2}}\left(1-{m_{h^0}^2 \over
m_t^2}\right)^2\left[2 \epsilon^4\{\left|{h_{23}^u \over h_{33}^u}\right|^2+
\left|{h_{32}^u \over
h_{33}^u}\right|^2\}\right]~.
\end{equation}
For $m_{h^0}= 100$ GeV, and with $|h_{23}^u/h_{33}^u| \simeq |h_{32}^u/h_{33}^u| \simeq
1$, the branching ratio is $Br(t\rightarrow ch^0) \simeq 1.1 \times 10^{-3}$.
With an integrated luminosity of $20 {\rm fb}^{-1}$, about $1.4 \times 10^{5}$~
$t\bar{t}$
pairs are expected at Tevatron running at $\sqrt{s} = 2$ TeV.  This would lead
to about 300 Higgs events from $t$ and $\bar{t}$ decays.  The invariant mass of
the $b\bar{b}$ jets will provide the Higgs signal.  The QCD background can be
brought under control by tagging the $t$ (or $\bar{t})$ by its decay into $W +b$.
This process could be useful to discover a Higgs boson of mass as large as about
150 GeV at the Tevatron (at which point the kinematic suppression becomes significant).
This reach can be as large as 170 GeV at LHC.

For $m_{h^0}$ between $m_t$ and $2m_t$, there is another way to look for the Higgs boson
at hadron colliders.  The cross section for $h^0$ production via $gg$ fusion is
about 15 pb for $m_{h^0} = 200$ GeV at the LHC running at $\sqrt{s}=14$ TeV.
The Higgs will decay dominantly into
$W$ pairs, but the branching fraction into $\bar{t}c+\bar{c}t$ is not negligible.
For $h_{23}^u \simeq h_{32}^u \simeq 1$, the branching ratio, allowing for the
kinematic suppression, is found to be $Br(h^0\rightarrow
\bar{t}c+\bar{c}t) \simeq 1\times 10^{-3}$.  With $100 {\rm fb}^{-1}$
of data, this will result in 1500 $\bar{t}c + \bar{c} t$ events.
The signal will be thus a $b$--jet, a charm and a $W$.
It might be possible
to reconstruct the invariant mass of the Higgs boson by studying the leptonic
decay of the $W$ from the top quark.  (Although there is a neutrino involved, its
four--momentum can be reconstructed by measuring the charged lepton momentum, up to
a two--fold ambiguity.)  The background from SM single top production can be
substantially reduced by the invariant mass requirement.  Two jet $+W$ production
(where a jet is misidentified as a $b$) and $\bar{b}bW$ where one $b$ is missed
are other dominant backgrounds.  The presence  of a top quark in the signal but
not in these background events can be utilized to provide further cuts\footnote{We thank
John Conway for very helpful discussions on this signal.}.

5.  Since the scale of flavor physics is identified to be $M \approx
(1-2)$ TeV from fermion masses and mixings, new phenomena associated with
flavor physics will show up at experiments performed with energies
greater than $(1-2)$ TeV.  This will happen at the LHC.  One concrete example
is the unraveling of the effective vertices in Eq. (1).  For example, at
$s \sim M^2$, the process $p p \rightarrow \bar{b}b (3h^0)$ will proceed without
much suppression in the coupling.
The relevant dimensionful coupling
is $h_{33}^d/M^2 \simeq m_b/v^3$.  For $m_{h^0} \sim 100$ GeV, we estimate
the cross section for this process at LHC to be in the fb range.
The signature will be quiet dramatic: 8 $b$--jets with
3 pairs adding to to the same invariant Higgs mass.

Before concluding we wish to remark briefly on possible ways of inducing the
effective Yukawa Lagrangian on Eq. (1).  It is quite conceivable that it has
a dynamical origin \cite{nambu}.  In order to show that Eq. (1) can be derived from
perturbative physics at the TeV scale we have constructed models which
should serve as proof of existence.  Suppose there are vector--like
fermions at the TeV scale in doublet and singlet representations.  They could
have flavor--symmetric Yukawa couplings to the SM fermions and the SM Higgs
doublet.  The masses of these vector fermions break the flavor symmetry
softly (or spontaneously).  Once these vector fermions are integrated out,
the effective Lagrangian of Eq. (1) will result.  Since the extra matter is
vector--like under the SM, no significant departures from the SM predictions
are expected in the precision
electroweak observables.

In summary, we have suggested a novel scenario to explain the observed
hierarchy of the fermion masses and mixings.  By using successive higher dimensional
operators involving the SM Higgs doublet field we obtained a
good fit to the fermion masses and mixings without putting small numbers
into the theory.  We pointed out several new features associated with the
Higgs boson that are predicted by this scheme.  While low energy constraints
are well satisfied, there is one prediction in $D^0-\bar{D^0}$ mixing that
will test the model.  The flavor changing $\bar{t}ch^0$ coupling, as large in
strength as the flavor conserving $\bar{b}hh^0$ coupling, leads to new possibilities
to discover the Higgs boson at the upgraded Tevatron and the LHC.

We wish to thank Mike Berger, John Conway and Sally Dawson for discussions.
This work is supported in part by  the Department of Energy Grant No. FG03-98ER41076
and by funds provided by the Oklahoma State University.


\begin{thebibliography}{99}

\bibitem{other}
For attempts to  address the fermion mass hierarchy problem see: \\
C.D. Froggatt and H.B. Nielsen, Nucl. Phys. {\bf B147}, 277 (1979);\\
S.M. Barr, Phys. Rev. {\bf D21}, 1424 (1980);\\
H. Georgi, A. Manohar and A. Nelson, Phys. Lett. {\bf 126B}, 169 (1983);\\
B.S. Balakrishna, A.L. Kagan and R.N. Mohapatra, Phys. Lett. {\bf 205B},
345 (1988);\\
K.S. Babu and R.N. Mohapatra, Phys. Rev. Lett. {\bf 64}, 2747 (1990);\\
Z. Berezhiani and R. Rattazzi, Nucl.Phys. {\bf B407}, 249 (1993);\\
Y. Nir and N. Seiberg, Phys. Lett. {\bf B309}, 337 (1993);\\
J. Elwood, N. Irges and P. Ramond, Phys. Lett. {\bf B413}, 322 (1997);\\
J. Ellis, S. Lola and G. Ross, Nucl. Phys. {\bf B526}, 115 (1998);\\
S. Barbieri, L. Hall and A. Romanio, Nucl. Phys. {\bf B551}, 93 (1999).

\bibitem{muon}
For a review see: V. Barger, M. Berger, J. Gunion and T. Han, Phys. Rept.
{\bf 286}, 1 (1997).

\bibitem{gasser}
G. Gasser and H. Leutwyler, Phys. Rept. {\bf 87}, 77 (1982).

\bibitem{2higgs}
For discussions of FCNC processes in two Higgs doublet models, see eg:\\
T.P. Cheng and M. Sher, Phys. Rev. {\bf D35}, 3484 (1987);\\
Y.L. Wu and L. Wolfenstein, Phys. Rev. Lett. {\bf 73}, 1762 (1994);\\
A. Antaramian, L. Hall and A. Rasin, Phys. Rev. Lett. {\bf 69}, 1871 (1992);\\
L. Hall and S. Weinberg, Phys. Rev. {\bf D48}, 979 (1993).

\bibitem{soni}
See for e.g. D. Atwood, L. Reina,  and A. Soni,
Phys. Rev. {\bf D55}, 3156 (1997).

\bibitem{buras} For a review see:
S. Bosch et. al., hep-ph/9904408.

\bibitem{barr}
S. Barr and A. Zee, Phys. Rev. Lett. {\bf 65}, 21 (1990).

\bibitem{pdg}
Particle Data Table, C. Caso et. al., Eur. Phys. J. {\bf C3}, 1 (1998).

\bibitem{pakvasa}
For a review of FCNC in charm physics see: S. Pakvasa, hep-ph/9705397.

\bibitem{higgs}
For a review see:
 {\it The Higgs Hunters Guide}, J. Gunion, H. Haber,
G. Kane and S. Dawson, Addison-Wesley (1994).

\bibitem{dawson}
S. Dawson and L. Reina, hep-ph/9903360.

\bibitem{spira}
For recent reviews see:  M. Spira, hep-ph/9810289;\\
S. Dawson, hep-ph/9901280;\\
Summary talks by M. Carena and H. Haber, Run II Wrokshop at Fermilab, November (1998).

\bibitem{han}
For a recent discussion see:
T. Han and R-J. Zhang, Phys. Rev. Lett. {\bf 82}, 25 (1999).

\bibitem{nambu}
See for eg:
S. Dimopoulos and L. Susskind, Nucl. Phys. {\bf B155}, 237 (1979);\\
E. Eichten and K. Lane, Phys. Lett. {\bf 90B}, 125 (1980);\\
W. Bardeen, C. Hill and M. Lindner, Phys. Rev. {\bf D41}, 1647 (1990);\\
Y. Nambu, Nucl. Phys. {\bf A638}, 35 (1998);\\
C. Hill, Phys. Lett. {\bf B345}, 483 (1995).


\end{thebibliography}
\end{document}